  \documentclass[journal]{IEEEtran}

\usepackage[noadjust]{cite}

\ifCLASSINFOpdf
   \usepackage[pdftex]{graphicx}
\else
   \usepackage[dvips]{graphicx}
\fi
\usepackage[colorlinks,
            linkcolor=black,
            urlcolor=red,
            anchorcolor=black,
            citecolor=black
            ]{hyperref}
\usepackage[utf8]{inputenc}
\usepackage[braket]{qcircuit}
\usepackage{graphicx}
\usepackage{epstopdf}
\usepackage{dcolumn}
\usepackage{bm}
\usepackage{mathrsfs}
\usepackage{xcolor}
\usepackage{hyperref}
\usepackage{amsmath,amssymb}
\usepackage{makecell}
\usepackage{lineno}

\usepackage{fixltx2e}
\usepackage{url}

\begin{document}
\title{Partially Concatenated Calderbank-Shor-Steane Codes Achieving  the Quantum Gilbert-Varshamov Bound Asymptotically}
%

\author{Jihao~Fan,~\IEEEmembership{Member,~IEEE,}
        Jun~Li,~\IEEEmembership{Senior Member,~IEEE,}
         Ya~Wang, Yonghui~Li,~\IEEEmembership{Fellow,~IEEE,} Min-Hsiu Hsieh, ~\IEEEmembership{Senior Member,~IEEE,} and Jiangfeng~Du
        \thanks{This work was supported in part by the National Natural Science Foundation of China (No.   61802175), in part by the National Natural Science Foundation of China (No. 61872184),  in part by the Fundamental Research Funds for the Central Universities   (No. 30921013104), and in part by Future Network Grant of Provincial Education Board in Jiangsu.  \emph{(Corresponding
 authors: Jun Li; Min-Hsiu Hsieh.)} }
\thanks{J. Fan is with School of Cyber Science and Engineering, Nanjing University of Science and Technology, Nanjing 210094, China (e-mail: jihao.fan@outlook.com).}
\thanks{J. Li is with School of Electronic and Optical Engineering, Nanjing University of Science and Technology,  Nanjing  210094,  China (e-mail: jun.li@njust.edu.cn).  }
\thanks{Y. Wang and J. Du are with CAS Key Laboratory of Microscale Magnetic Resonance and Department of Modern Physics, University of Science and Technology of China, Hefei 230026, China (e-mail: \{ywustc, qcmr\}@ustc.edu.cn).}   
\thanks{Y. Li is  with
School of Electrical and Information Engineering, the University of Sydney, Sydney, NSW 2006, Australia (e-mail: yonghui.li@sydney.edu.au).}
 \thanks{M.H. Hsieh   is  with Quantum Computing Research Center, Hon Hai Research Institute, Taipei City 114, Taiwan (e-mail: min-hsiu.hsieh@foxconn.com).}
}

%
%

\markboth{IEEE TRANSACTIONS ON INFORMATION THEORY, August~2022}%
{Shell \MakeLowercase{\textit{et al.}}: Bare Demo of IEEEtran.cls for IEEE Journals}

%



\maketitle

\begin{abstract}
In this paper, we utilize a concatenation scheme to construct new families of quantum error correction codes achieving  the quantum Gilbert-Varshamov (GV) bound  asymptotically.  We  concatenate   alternant codes with any linear code achieving  the  classical  GV  bound
  to construct Calderbank-Shor-Steane (CSS) codes. We show that the  concatenated code  can achieve   the
quantum GV bound asymptotically and can approach the Hashing bound for
asymmetric Pauli channels.  By combing Steane's enlargement construction of CSS codes, we derive  a family of  enlarged stabilizer codes achieving  the quantum GV bound for enlarged CSS codes asymptotically.  As   applications, we derive  two families of fast encodable and decodable    CSS codes   with
parameters
$\mathscr{Q}_1=[[N,\Omega(\sqrt{N}),\Omega( \sqrt{N})]],$ and
$\mathscr{Q}_2=[[N,\Omega(N/\log N),\Omega(N/\log N)/\Omega(\log N)]].$
We show  that $\mathscr{Q}_1$ can be encoded
very efficiently by circuits of size     $O(N)$  and  depth $O(\sqrt{N})$.  For an input error syndrome,   $\mathscr{Q}_1$    can correct any adversarial  error of weight up to half the
minimum distance bound in $O(N)$ time.   $\mathscr{Q}_1$ can also  be decoded in parallel in $O(\sqrt{N})$ time   by using $O(\sqrt{N})$  classical  processors.
 For an input error syndrome, we proved that $\mathscr{Q}_2$ can correct  a linear number of ${X}$-errors with   high probability and an almost linear number  of  ${Z}$-errors     in $O(N )$ time. Moreover,  $\mathscr{Q}_2$ can be decoded in parallel in $O(\log(N))$ time   by using $O(N)$ classical processors.
\end{abstract}


%
\IEEEpeerreviewmaketitle

\section{Introduction}
\newcounter{conter1}
\setcounter{conter1}{0}
\newtheorem{theorem}[conter1]{Theorem}

\newtheorem{definitions}{Definition}
\newtheorem{lemmas}{Lemma}
\newtheorem{corollarys}{Corollary}
\newtheorem{examples}{Example}
\newtheorem{propositions}{Proposition}
%
%
%
%

Quantum systems are vulnerable to  the environment  noise induced by \emph{decoherence}, which is  one of the biggest obstacles in       quantum information processing. Similar to classical systems, one feasible solution is to exploit    quantum  error correction codes (QECCs) to  encode  the primitive quantum information into a larger quantum state. 
It is widely  believed that QECCs are  necessary  in realizing long-term quantum communications   and in building fault-tolerant   quantum computers \cite{deutsch2020harnessing,preskill2018quantum}. The construction and design of QECCs with excellent performance  is thus one  significant task.  QECCs can be constructed  from classical linear codes, e.g., by using the  {stabilizer} formalism \cite{calderbank1998quantum} or the  Calderbank-Shor-Steane (CSS) construction \cite{shor1995scheme,calderbank1996good}.
 But it is not    an easy task to  do that straightforwardly since an additional dual-containing constraint is needed.

 How to construct  asymptotically good QECCs with   positive rates and linear distances,
 is one of the key problems in    quantum coding theory \cite{calderbank1996good,christandl2020fault}. The quantum Gilbert-Varshamov  (GV) bound    is an important  lower bound that promises  the existence of such  good quantum codes \cite{calderbank1996good,ekert1996quantum, calderbank1997quantum,ashikhmin2000quantum2,ashikhmin2001nonbinary,matsumoto2017two}.
Compared with classical codes, it is  usually much more challenging to obtain  quantum codes that can   attain the quantum GV bound due to the  dual-containing constraint   \cite{calderbank1998quantum}. Therefore  very few types of QECCs can attain the quantum GV bound  up to now. In \cite{calderbank1996good}, it was shown that quantum codes derived from dual-containing codes can attain the quantum GV bound for CSS codes. In \cite{calderbank1996good},  random  codes were shown to attain the quantum GV bound for binary stabilizer codes. In \cite{ashikhmin2000quantum2}, stabilizer codes derived from self-orthogonal quaternary codes were shown to attain the quantum GV bound for binary stabilizer codes.
In   \cite{Li5165198}, asymptotically good  concatenated quantum codes (CQCs) were constructed. In \cite{ouyang2014concatenated}, it was shown that  CQCs can attain the quantum GV bound for general stabilizer codes asymptotically. A finite version of the quantum GV bound was given in \cite{feng2004finite}. In \cite{matsumoto2017two}, quantum GV bounds for asymmetric quantum codes (AQCs) were given. All the known asymptotic results about quantum GV bound rely  on some random code arguments  which mean  that there is little structure in them.

It should be noted that, a unique phenomenon  in quantum coding theory, called \emph{error degeneracy}, exists, which makes degenerate codes   more powerful  than  nondegenerate ones \cite{PhysRevA57830,FanPNAS2022}.    In an extreme case, it is possible to construct asymptotically good quantum codes from asymptotically bad classical codes. We refer to such codes as \emph{extremely} degenerate codes. For example,
  the qLDPC conjecture asks   whether there exist families of
  asymptotically good quantum low-density parity-check (qLDPC) codes   \cite{evra2020decodable,hastings2020fiber,breuckmann2021quantum}.
Recently, it was shown that asymptotically good qLDPC codes exist with high probabilities   \cite{panteleev2020quantum,panteleev2021asymptotically} and  the resultant qLDPC codes are extremely degenerate. The previous results about quantum GV bounds usually rely on asymptotically good  classical linear codes. It was unknown   whether there exist  extremely degenerate codes derived from bad classical codes  can attain  the   quantum GV bound.

In \cite{Steane796388},    an enlargement   of the CSS construction was  proposed to construct more efficient stabilizer codes than the standard CSS codes. In \cite{Ling5508639}, the enlargement was  generalized to   nonbinary situations. Several asymptotic bounds for enlarged CSS codes were also given in  \cite{Steane796388,Ling5508639}. Then in  \cite{ashikhmin2001asymptotically,matsumoto2002improvement,chen2001asymptotically}, asymptotically good binary quantum algebraic geometry (AG) codes were constructed by using   enlargement of the CSS construction.  In \cite{Hamada2008concatenated}, asymptotically good   concatenated quantum AG codes  were constructed by using the CSS construction and its enlargement, and these codes can  be decoded efficiently in polynomial time.
 However, whether binary enlarged CSS codes  can attain the corresponding quantum GV bound is unknown.     On the other hand, it should be noted that quantum AG codes over  sufficiently  large field size   can even beat the quantum GV bound \cite{feng2006asymptotic,Niehage2007,wang2010asymmetric}. However, for  codes over small field size, e.g., binary codes,  quantum GV bound is much better than   asymptotic bounds from quantum AG codes.

In addition to the construction of   QECCs with good parameters, practical QECCs need to equip with both efficient encoders and decoders. If the syndrome decoding is slower than the error accumulations,  additional noise will be introduced during the syndrome decoding \cite{holmes2020nisq+}. {However, error degeneracy does not simplify the syndrome decoding of QECCs \cite{hsieh2011np}.  In contrast,
it makes most decoding methods that are efficient for classical codes  fail to fully decode QECCs,    e.g., the belief-propagation (BP)   algorithm \cite{Poulin2008}. How to utilize
error degeneracy to correct more errors is an attractive
and crucial issue, and its importance, in our opinion, is
no less than the construction of good QECCs.
  }

On the other hand, the encoding of QECCs is   also     significant        but it has received much less attention compared to  the decoding.  Noise not only occurs in the transmission and decoding, but also in the encoding circuit \cite{christandl2020fault}. Large-scale encoding circuit and deep encoding depth  make the risk of qubits suffering   from  faulty gates  increase rapidly. Consequently, the qubits might be corrupted even before the transmission and the computation. The encoding circuit of  stabilizer codes of length $N$   with $g$ stabilizer generators generally  has   $O(gN)$ gates   and  $O(N)$ depth \cite{cleve1997efficient}. But in  many practical applications, the encoding circuits need to be with linear size and sublinear depth \cite{dennis2002topological,higgott2020optimal}.

In this work, we  combine  classical   alternant codes with  some   other linear codes  to construct QECCs achieving  the quantum GV bound asymptotically. We use classical alternant codes as the outer code  and use any linear code achieving  the classical GV bound as the inner code to do the concatenation. We call the constructed   codes as  partially concatenated CSS (PC-CSS) codes since we only introduce   concatenation in the $X$-stabilizers (or the $Z$-stabilizers). We show that   PC-CSS codes can  attain    quantum GV bounds for CSS codes and asymmetric CSS codes asymptotically. Here, our scheme can use any linear code  achieving  the classical GV bound to do the concatenation. Then PC-CSS codes can   achieve  the quantum GV bound asymptotically.
In particular, if   we use low-density parity-check (LDPC) codes as the component codes, we show that error degeneracy can greatly improve the minimum distance of  PC-CSS codes. In such case,
the   minimum distance of PC-CSS codes without considering error degeneracy is constant to the block length.
Yet,   degenerate PC-CSS codes can achieve  the quantum GV bound whose minimum distance is linear to
the block length.  Thus PC-CSS codes are extremely degenerate and error degeneracy has large advantages in constructing asymptotically good quantum codes. It should be noted that PC-CSS codes are not qLDPC codes   since only $Z$-stabilizers of PC-CSS codes  can satisfy the qLDPC constraints \cite{hastings2021quantum}.  It is known that  asymptotically good qLDPC codes   has been shown to exist with high probabilities \cite{panteleev2020quantum,panteleev2021asymptotically}.   However, the asymptotic bound of qLDPC codes in \cite{panteleev2020quantum,panteleev2021asymptotically} is much weaker than quantum GV bound. We show that PC-CSS codes can not only attain the quantum GV bound but also can approach  the capacity of Pauli  channels with large asymmetries.

\begin{table}[t!]
\centering
 \caption{Summary of Asymptotic  Bounds for  Different  Quantum Codes. Let $p$  be a prime and let  $m>0$ be an integer. Denote by $q=p^m$. For a stabilizer code with parameters $Q=[[n,k,d]]_q$, we denote by $R=k/n$ and $\delta=d/n$. For a nonstabilizer   code with parameters $Q=((n,K,d))_q$, we denote by $R=(\log_q K)/n$ and $\delta=d/n$. For an AQC with parameters $Q=[[n,k,d_X/d_Z]]$, we denote by $R=k/n$,  $\delta_X=d_X/n$, and $\delta_Z=d_Z/n$.}
\begin{tabular}{|c|c|c|c|c|c|c|}
\hline
Nos.   &References   &   \makecell[c]{Asymptotic Bounds}        \\
\hline
1&\cite{calderbank1998quantum}&  $R\geq 1-2H_2(\delta)-O(1)$  \\
\hline
2&  \cite{calderbank1996good,ashikhmin2000quantum2} &$R\geq 1-2H_4(\delta)-O(1)$   \\
\hline
3&  \cite{ouyang2014concatenated} &\makecell[c]{$R\geq 1-2H_{q^2}(\delta)-O(1)$ } \\
\hline
4&  \cite{ashikhmin2001asymptotically}&\makecell[c]{$R= 1- \frac{1}{2^{m-1}-1}-\frac{10}{3}m\delta$, \\$\delta\in[\delta_m,\delta_{m-1}]$, $m=3,4,5,\ldots,$ \\$\delta_2=1/18$, $\delta_3=3/56$, and \\ $\delta_m=\frac{3}{5}\frac{2^{m-2}}{(2^m-1)((2^{m-1}-1))}$,\\ $m=4,5,6,\ldots $ } \\
\hline
5&  \cite{matsumoto2002improvement}    &\makecell[c]{$R=1-\frac{10}{3}m\delta-\frac{2}{2^m-1}$, \\
    $0< \delta\leq\frac{1}{2m}(\frac{1}{2}-\frac{1}{2^m-1})$}  \\
\hline
6&  \cite{chen2001asymptotically}  &\makecell[c]{$R= 3t(\delta_t-\delta)$, \\ $\delta_t=\frac{2}{3}\frac{2^t-3}{3(2t+1)(2^t-1)}$,\\$t\geq 3,$ $0<\delta<\delta_t$}
      \\
\hline
7&  \cite{Li5165198}     &\makecell[c]{$R\geq \frac{1}{3}(1- 4\delta/a ) $,\\$a=H^{-1}_4(1/6)$ }
      \\
\hline
8&  \cite{Hamada2008concatenated}    &\makecell[c]{$R\geq   \frac{t}{t+1}-\frac{2t}{(t+1)(q^t-1)}  -2t\delta $,\\ $t>0$ is an integer}
      \\
\hline
 9&  \cite{Hamada2008concatenated}     &\makecell[c]{$R\geq  \frac{k_0}{n_0}(1-2\gamma-\frac{(2q+1)n_0}{(q+1)d_0}\delta), $\\ $0\leq \gamma\leq (q^{k_0/2}-2)^{-1}$,\\$[[n_0,k_0,d_0]]_q$ is a quantum code}
     \\
\hline
   10 &  \cite{feng2006asymptotic,Niehage2007}   &\makecell[c]{$R\geq 1-2\delta-\frac{2}{\sqrt{q}-1} $,\\$q$ is a square }
     \\
\hline
  11 &  \cite{feng2006asymptotic}    &\makecell[c]{$R\geq 1-2\delta-\frac{2}{\sqrt{q}-1}+\log_q(1+\frac{1}{q^3}) $,\\$q$ is a square }
     \\
\hline
 12 &    \cite{panteleev2021asymptotically}  &\makecell[c]{$0<R<1$, $0<\delta<\frac{1}{2c^{7/2}}$, \\ $c>1$ is an integer  }
     \\
\hline
 13&  \cite{matsumoto2017two}    &\makecell[c]{$R\geq  1-H_q(\delta_X)-H_q(\delta_Z) -O(1)$,\\ $0\leq\delta_X\leq1-1/q$,\\$0\leq\delta_Z\leq1-1/q$ }
     \\
     \hline
 14&  \cite{wang2010asymmetric}   &\makecell[c]{$R\geq 1- \delta_X-\delta_Z-\frac{2}{\sqrt{q}-1} +\log_q(1+\frac{1}{q^3})$,\\$q$ is a square }
      \\
\hline
15& Theorem \ref{Degenerate QSCPC}  &  $R\geq 1-2H_q(\delta)$  \\
\hline
16&  Corollary \ref{asymmetricenlarge}    &\makecell[c]{$R\geq  1-H_q(\delta_X)-H_q(\delta_X)-O(1) $,\\ $0\leq\delta_X\leq1-1/q$,\\$0\leq\delta_Z\leq1-1/q$ }
      \\
\hline
 17 & Theorem  \ref{enlargedCSSGV}  &\makecell[c]{$R\geq1- H_{q}(\delta )-H_{q}(\frac{q}{q+1}\delta  )-O(1) $}
     \\
\hline
\end{tabular}
\label{ComparisonsAsymptoticBounds}
\end{table}

 As a special case, we construct a family of asymmetric PC-CSS codes with parameters $
 \mathscr{Q}=[[N,\Omega(N/n_0),d_{Z}\geq\Omega(N/n_0)/d_{X}\geq n_0]]
 $
  by using expander codes as one of the component codes, where   $n_0>1$ and $N>1$ are   integers and   $n_0|N$.
Let $n_0=\Theta(\sqrt{N})$,     we can derive
  a family of quantum codes with parameters
$
  \mathscr{Q}_1=[[N,\Omega(\sqrt{N}),\Omega(\sqrt{N})]].
$
For an input error syndrome, $ \mathscr{Q}_1$ can correct all errors of weight smaller than  half   the minimum distance bound   in $O(N)$ time.     $\mathscr{Q}_1$ can also be decoded  in parallel in $O(\sqrt{N})$ time  by using $O(\sqrt{N})$  classical  processors. We  show that $\mathscr{Q}_1$ can be fast encoded  by a circuit of size $O(N)$ and depth $O(\sqrt{N})$. Moreover, $\mathscr{Q}_1$ can correct a random $Z$-error   with very high probability  provided $p_{z}<25\%$.

If   let  $n_0=\Theta(\log N)$, then we can derive a family of AQCs with parameters
$
 \mathscr{Q}_2=[[N,\Omega(N/\log N),\Omega(N/\log N)/\Omega(\log N)]].
$
For an input error syndrome, $\mathscr{Q}_2 $ can correct any  random   ${X}$-error  with   high probability and correct an almost linear number  of  ${Z}$-errors  in $O(N)$ time.  Further, we show that $\mathscr{Q}_2$ can be decoded in parallel in $O(\log(N))$ time   by using $O(N)$ classical processors. While fault-tolerant quantum computation   \cite{gottesman2014fault} and communication   \cite{muralidharan2014ultrafast,christandl2020fault} prefer fast  encodale and decodable codings,  $\mathscr{Q}_1$ and $\mathscr{Q}_2$ are thus practical and beneficial for the future use.

We further combine our concatenation  scheme with  the enlarged   CSS construction in \cite{Steane796388,Ling5508639}. We show that  the enlarged quantum codes can attain the quantum GV bound for enlarged CSS codes.  In Table \ref{ComparisonsAsymptoticBounds}, we give a summary of asymptotic  bounds for  different  quantum codes.

This paper is organized as follows. We begin in Section \ref{Preliminaries} with Preliminaries on   classical linear codes and quantum stabilizer codes. In Section  \ref{enlargementAlternant}, we present the partial concatenation  construction of CSS codes and show that PC-CSS codes can attain the quantum GV bound for CSS codes and AQCs asymptotically. In Section \ref{FastDecEnc},
 we give a family of fast encodable and decodable PC-CSS codes. In Section \ref{enlargementSteane}, we combine our concatenation  scheme with  Steane's   enlarged   CSS construction. We show that the enlarged quantum codes can attain the quantum GV bound for enlarged CSS codes asymptotically. The discussions and conclusions are given in Section \ref{DisCon}.

\section{Preliminaries}
\label{Preliminaries}
  We   shortly introduce    some needed knowledge about classical linear codes and quantum stabilizer codes, for more details, see the literatures, e.g., \cite{macwilliams1981theory,calderbank1996good,calderbank1998quantum,ketkar2006nonbinary,sarvepalli2009asymmetric}. In this paper,  we usually neglect
the index in the code parameters for binary linear codes and
binary stabilizer codes  provided there is no ambiguity.

\subsubsection{Classical Alternant Codes}  Let $p\geq2$ be a prime and denote by $GF(p)$   a prime field. Let $q=p^{m_0}$ be a power of $p$,   where $m_0>0$ is an integer. Let $GF(q)$ be the Galois field of size $q$ and denote by    $ GF(q^m)$     a field extension  of
  $GF(q)$, where  $m>0$ is an integer. Define the trace operation from  $GF(q)$ to $GF(p)$ as $\textrm{Tr}(\alpha)=\sum_{i=0}^{m_0-1}\alpha^{p^i}$, where $\alpha\in GF(q)$.  We review some known results about classical alternant codes  which will be used in the construction of quantum codes.
  Let $\alpha_1, \alpha_2, \ldots, \alpha_n$ be $n$ distinct elements of $GF(q^m)$ and let $v_1, v_2, \ldots, v_n$ be $n$ nonzero elements of $GF(q^m)$, where $2\leq n\leq q^m$. Denote by $\mathbf{a}=(\alpha_1, \alpha_2, \ldots, \alpha_n)$ and  $\mathbf{v}=(v_1, v_2, \ldots, v_n)$.  Denote   the ring of polynomials with coefficients in  finite field $GF(q^m)$   by $ GF(q^m)[x]$.
For any polynomial $F(x)=\sum_{i=0}^{l}c_ix^i\in GF(q^m)[x]$, the degree of $F(x)$   is denoted by $\deg F(x)=l$. For any $1\leq k\leq n-1$,
the generalized Reed-Solomon (GRS) code $\textrm{GRS}_k(\mathbf{a},\mathbf{v})$ is defined by
\begin{eqnarray}
\label{definition of GRS codes}
\nonumber
\textrm{GRS}_k(\mathbf{a},\mathbf{v})\equiv\big\{(v_1F(\alpha_1), v_2F(\alpha_2), \ldots, v_nF(\alpha_n))\ |\\\
 F(x)\in GF(q^m)[x],\ \deg F(x)<k \big\}.
\end{eqnarray}

$\nrightarrow$
The parameters of $\textrm{GRS}_k(\mathbf{a},\mathbf{v})$ are given by
$[n,k,n-k+1]_{q^m},$ and  GRS codes are maximum-distance-separable (MDS) codes which can attain the Singleton bound in \cite{macwilliams1981theory}. Let $r=n-k$.
 The dual of a GRS code is also a
GRS code, i.e., $\textrm{GRS}_k(\mathbf{a},\mathbf{v})^\bot=\textrm{GRS}_{r}(\mathbf{a},\mathbf{y})$,
where  $\mathbf{y}=(y_1, y_2, \ldots, y_n)$ and $y_i\cdot v_i=1/\prod_{j\neq i}(\alpha_i-\alpha_j)$, for $1\leq i\leq n$. The parity-check
matrix of  $\textrm{GRS}_k(\mathbf{a}, \mathbf{v})$ is given by
\begin{equation}
\label{parity-check matrix of alternant codes}
H_{\textrm{GRS}_k(\mathbf{a},\mathbf{v})} =
\left(
\begin{array}{cccc}
y_1&y_2&\cdots&y_n\\
\alpha_1y_1&\alpha_2y_2&\cdots&\alpha_ny_n\\
\vdots&\vdots&\vdots&\vdots\\
\alpha_1^{r-1}y_1&\alpha_2^{r-1}y_2&\cdots&\alpha_n^{r-1}y_n
\end{array}
\right).
\end{equation}

For some given GRS code $\textrm{GRS}_k(\mathbf{a}, \mathbf{v})$ over $GF(q^m)$, the alternant code   over $GF(q) $ denoted by $\mathcal{A}_r(\mathbf{a}, \mathbf{y})$, is defined    as the subfield subcode of $\textrm{GRS}_k(\mathbf{a}, \mathbf{v})$, i.e.,
\begin{equation}
\label{DefAlternantCodes}
\mathcal{A}_r(\mathbf{a}, \mathbf{y})=\textrm{GRS}_k(\mathbf{a}, \mathbf{v})| GF(q) .
\end{equation}
   Since $\mathcal{A}_r(\mathbf{a}, \mathbf{y})$ consists of all codewords of $ \textrm{GRS}_k(\mathbf{a}, \mathbf{v})$ over $GF(q)$, the parity check matrix of $\mathcal{A}_r(\mathbf{a}, \mathbf{y})$ is given by that of   $ \textrm{GRS}_k(\mathbf{a}, \mathbf{v})$ in
 (\ref{parity-check matrix of alternant codes}).
The integer $r=n-k$ is called the degree of the alternant code $\mathcal{A}_r(\mathbf{a}, \mathbf{y})$. In addition,  $\mathbf{a}$ is called the support of $\mathcal{A}_r(\mathbf{a}, \mathbf{y})$, and $\mathbf{y}$ is called a multiplier of $\mathcal{A}_r(\mathbf{a}, \mathbf{y})$. For the dimension and minimum distance of alternant codes, there is the following result:
\begin{lemmas}{\cite{macwilliams1981theory}}
\label{dimanddisofalternantcodes}
Let $ \mathscr{C}=\mathcal{A}_r(\mathbf{a}, \mathbf{y})  =[n ,k_\mathscr{C},d_\mathscr{C}]_q$ be an alternant code  defined  in (\ref{DefAlternantCodes}). The dimension and minimum distance of   $ \mathscr{C}$ can be given by $k_\mathscr{C}\geq n-mr$ and $d_\mathscr{C}\geq r+1$, respectively.
\end{lemmas}

Let $u$  be any nonzero vector over $GF(q)$. Let  $\mathbf{a}$ be a fixed vector and let     $\mathbf{v}$ be a  varying vector. According to \cite[Chap. 12, Theorem 3]{macwilliams1981theory}, we know that the number of GRS codes
$\textrm{GRS}_k(\mathbf{a},\mathbf{v})$   containing $u$ is at most $(q^m-1)^{k}$.  Therefore, the number of alternant codes $ \mathcal{A}_r(\mathbf{a}, \mathbf{y})=\textrm{GRS}_k(\mathbf{a}, \mathbf{v})| GF(q)$ containing $u$ is also at most $(q^m-1)^{k}$. Here, it should be noted that  the estimate of the number of alternant  codes that contain $u$ does not relate to its Hamming weight.
\begin{lemmas}{\cite{macwilliams1981theory}}
\label{numberofalternantcodes}
 For  any nonzero vector $u\in GF(q)^n$, the number of alternant codes $ \mathcal{A}_r(\mathbf{a}, \mathbf{y}) $ in (\ref{DefAlternantCodes}) that  contain $u$ is   at most $(q^m-1)^{n-r}$.
\end{lemmas}

 On the other hand, the total number of alternant codes $\mathcal{A}_r(\mathbf{a}, \mathbf{y})$ is equal to the number of  choices for different $\mathbf{y}$ (or $\mathbf{v}$), which is given by $(q^m-1)^n$ \cite{macwilliams1981theory}.
 Thus alternant codes form a large family of linear codes and  include many important subclasses, e.g.,   Goppa codes and Bose-Chaudhuri-Hocquenghem  codes. Further, alternant  codes play a significant role in the McEliece cryptosystem  \cite{faugere2015folding}. Moreover, alternant codes can attain the classical GV bound asymptotically \cite{macwilliams1981theory}. In the next section, we will show that alternant codes can be used to construct  concatenated CSS codes achieving  the quantum GV bound asymptotically.

For integers $n,\lambda$, denote by $\textrm{Vol}_q(n,\lambda)=\sum_{i=0}^{\lambda}\binom{n}{i}(q-1)^i$ the Hamming ball of radius $\lambda$. Denote by $H_q(x)=x\log_q(q-1)-x\log_qx-(1-x)\log_q(1-x)$   the $q$-ary entropy function. There exists the following important estimate of the Hamming ball.
\begin{lemmas}{\cite{Guruswami2010}}
\label{HammingBallEst}
Let $q$ be a power of a prime $p\geq2$. For an integer  $n>1$ and $0\leq\ell\leq1-1/q$,
\begin{equation}
q^{(H_q(\ell)-O(1))n}\leq \textrm{Vol}_q(n,\ell n)\leq q^{H_q(\ell)n}
\end{equation}
\end{lemmas}

\subsubsection{Quantum Stabilizer Codes}  Let $\xi=\exp(2\pi i/p)$ be a primitive
$p$th root of unity and let $u,v\in GF(q)$. Denote the unitary operators $X(u)$ and $Z(v)$ on $\mathbb{C}^q$ by
$
X(u)|a\rangle = |a+u\rangle$ and $Z(v)|a\rangle=\xi^{\textrm{Tr}(va)}|a\rangle
$, respectively. Denote by the finite   group
\begin{equation}G_{n,q}=
\left\{
   \begin{aligned}
    \langle X(\textrm{u}),Z(\textrm{v})|\textrm{u},\textrm{v}\in GF(q)^n\rangle,& &\textrm{if}&\
    p\  \textrm{is odd},  \\
 \langle iI, X(\textrm{u}),Z(\textrm{v})|\textrm{u},\textrm{v}\in GF(q)^n\rangle,& &\textrm{if}&\
  p\  \textrm{is even}.
 \end{aligned}
\right.
\end{equation}

A stabilizer code $Q$ is a $q^k$-dimensional   subspace of the    Hilbert space $ \mathbb{H}_n= \mathbb{C}^{q^n}$, i.e.,
\begin{equation}
Q= \{ |\varphi\rangle\in \mathbb{H}_n\ |\ E|\varphi\rangle=|\varphi\rangle,\ \forall\ E\in \mathcal{S}\},
\end{equation}
where $\mathcal{S}=\langle S_1,\ldots,S_{n-k}\rangle$ with stabilizer generators $S_i$($1\leq i\leq n-k$) is an Abelian     subgroup of $G_{n,q}$. $\mathcal{S}$ is called the   stabilizer group. By measuring the eigenvalues  of the stabilizer generators,  the syndrome  of   errors can be revealed. An error $E\in G_{n,q}$ is detectable if it anticommutes  with some stabilizer generator  $S_i(1\leq i\leq n-k)$.  A detectable error  results in a nonzero syndrome, otherwise  the syndrome is zero and the error is undetectable.
The minimum distance $d$ of   code $Q$ is the minimum weight of an
undetectable  error $E\in G_{n,q}$, which does not belong to the stabilizer group.  We denote by $Q=[[n,k,d]]_q$. The stabilizer code $Q$ is called nondegenerate if the stabilizer group $\mathcal{S}$ does not contain  a nontrivial element whose weight is less than $d$, otherwise it is degenerate. 

The CSS construction  in    \cite{calderbank1996good,steane1996simple} presents     a direct way to   construct stabilizer codes from two classical linear    codes  that satisfy the dual-containing  relationship.
\begin{lemmas}[{\cite{calderbank1996good,steane1996simple}}]
\label{CSS Constructions}
Let $C_1$ and $C_2$ be two $q$-ary
linear codes with parameters
$[n,k_1,d_1]_q$ and $[n,k_2,d_2]_q$, respectively. If $C_2^{\bot}\subseteq C_1$, then there exists a stabilizer code $Q$ with parameters $  [[n,k_1+k_2-n,d_Q]]_q,$ where
\begin{equation}
d_Q=\min\{\text{wt}_H(C_1\backslash C_2^{\bot}),\text{wt}_H(C_2\backslash C_1^{\bot})\}.
\end{equation}
If $d_Q=\min\{d_1,d_2\}$, then $Q$ is nondegenerate, otherwise it is degenerate.
\end{lemmas}

The enlargement of CSS construction in \cite{Steane796388,Ling5508639} can lead to more efficient stabilizer codes than the standard CSS construction.

\begin{lemmas}[\cite{Steane796388,Ling5508639}]
\label{enlargedCSSConstruction}
Let $C =[n,k,d]_q$ and $D=[n,k',d']_q$ be two  classical linear codes over $GF(q)$ such that $C^\bot\subseteq C\subseteq D$ and $k'>k+1$. Then there exists a stabilizer code with parameters
$
Q_E=[[n,k+k'-n,d_{Q_E}]]_q,
$
 where $d_{Q_E}=\min\{d,\lceil\frac{q+1}{q}d'\rceil\}$.
\end{lemmas}

In most quantum channels, the dephasing errors ($Z$-errors) usually happen much more frequently than the amplitude errors ($X$-errors). Asymmetric quantum codes (AQCs) are thus proposed to deal with such biased quantum noise  \cite{ioffe2007asymmetric,wang2010asymmetric,Fan2021asymmetric}. Moreover, the CSS construction can be used to derive AQCs with two classical linear codes, one for correcting   $Z$-errors and
the other  for correcting   $X $-errors.

\begin{lemmas}[{\cite{wang2010asymmetric}}]
\label{AQC Constructions}
Let $C_X$ and $C_Z$ be two $q$-ary
linear codes with parameters
$[n,k_1,d_1]_q$ and $[n,k_2,d_2]_q$, respectively. If   $C_Z^{\bot}\subseteq C_X$, then there exists a $ {Q_A}=[[n,k_1+k_2-n,d_Z/ d_X]]_q $ AQC, where
\begin{eqnarray}
d_Z&=&\max\{\text{wt}_H(C_X\backslash C_Z^{\bot}),\text{wt}_H(C_Z\backslash C_X^{\bot})\},\\
d_X&=&\min\{\text{wt}_H(C_X\backslash C_Z^{\bot}),\text{wt}_H(C_Z\backslash C_X^{\bot})\}.
\end{eqnarray}
If $\text{wt}_H(C_X\backslash C_Z^{\bot})=d_1$ and $ \text{wt}_H(C_Z\backslash C_X^{\bot})=d_2$, then $Q_A$ is nondegenerate, otherwise it is degenerate.
\end{lemmas}

\section{Asymptotic  Goodness of Partially Concatenated  Calderbank-Shor-Steane Codes}
\label{enlargementAlternant}
Denote by $C_1=[n_1,k_1,d_1]_q$   a  linear code with the parity check   and   generator matrices given by $H_1$ and $G_1$, respectively.   Let $C_2=[k_1,k_2,d_2]_q$ be  another linear code  with the parity check   and   generator matrices given by $H_2$ and $G_2$, respectively. We use $C_1$ as the inner code and $C_2$ as the outer code, and we concatenate the generator matrix of $C_1$ with the parity-check matrix of $C_2$ as follows:
\begin{equation}
\label{QSCPCHX}
\mathscr{H}_{{X}} = H_{2}G_{1}.
\end{equation}
Denote the null space of $\mathscr{H}_X
$ by $\mathscr{C}_{X}$ and we have $\mathscr{C}_{X}=[n_1,k_X]_q$, where $k_X=n_1-k_1+k_2$.   Denote by
\begin{equation}
\label{QSCPCHZ}
\mathscr{H}_{{Z}} = H_1\ \textrm{and} \ \mathscr{C}_{Z}=C_1.
\end{equation}
It is    easy to see that $\mathscr{H}_{{X}}\mathscr{H}_{{Z}}^T=0$ and
 $\mathscr{C}_{X}^{\bot}\subseteq  \mathscr{C}_{Z}$. Thus we can derive  a partially concatenated CSS (PC-CSS) code with parameters  $\mathscr{Q}  = [[n_1,k_2]]_q$ by using the CSS construction.
  We  show that   PC-CSS codes    can attain the quantum GV bound asymptotically.
\begin{theorem}
\label{Degenerate QSCPC}
There exist PC-CSS codes  with parameters $\mathscr{Q}=[[n_\mathscr{Q},k_\mathscr{Q},d_\mathscr{Q}]]_q$ achieving  the quantum GV bound asymptotically. $\mathscr{Q}$ can achieve a quantum rate
\begin{equation}
  \frac{k_\mathscr{Q}}{n_\mathscr{Q}}\geq1-2H_{q}(\delta_\mathscr{Q})-O(1)
\end{equation}
for any block length $n_\mathscr{Q}\rightarrow\infty$, where $\delta_\mathscr{Q}=d_\mathscr{Q}/n_\mathscr{Q}$ is the relative minimum distance.
\end{theorem}

\begin{IEEEproof}
 Let $C_1=[n_1,k_1,d_1]_q$ be any classical linear code. Let $C_2=[k_1,k_2,d_2]_q$ be an alternant code. We construct a pair of dual-containing codes $\mathscr{C}_{X}^{\bot}\subseteq  \mathscr{C}_{Z}$  according to (\ref{QSCPCHX}) and (\ref{QSCPCHZ}). By using the CSS construction, we can derive a PC-CSS code $\mathscr{Q}= [[n_1,k_2,d_{\mathscr{Q}}]]_q$ with
\begin{eqnarray}
\label{distanceQ1}
\nonumber
 d_{\mathscr{Q}}&=&\min \{\text {wt}_H(c)|c \in ( \mathscr{C}_Z\backslash \mathscr{C}_{X}^{\bot}) \cup(\mathscr{C}_{X}\backslash \mathscr{C}_Z^{\bot}) \}\\
 &=&\min\{ \omega_1,\omega_2\},
\end{eqnarray}
where $\omega_1=\min\{\text {wt}_H(c)|c \in \mathscr{C}_Z\backslash \mathscr{C}_{X}^{\bot}\}\geq  d_{1}$, and  $\omega_2=\min\{\text {wt}_H(c)|c\in \mathscr{C}_{X}\backslash \mathscr{C}_Z^{\bot} \}$.  Since $ \mathscr{C}_Z=C_1$ can be chosen arbitrarily,   we let $C_1$ be
an  asymptotically good linear code which can attain the classical GV bound, i.e.,
\begin{equation}
\label{GVC1}
 \frac{k_1}{n_1} \geq 1-H_q(\delta_1)-O(1),
\end{equation}
where $\delta_1=d_{1}/n_{1}$ is the relative minimum distance.   Next  we need to determine  $\omega_2$. Since we    know nothing about the  distance of the dual  code $C_1^{\bot}$, there
maybe some (very) low weight vectors coming from $C_1^{\bot}$. If we do not consider error degeneracy, the minimum distance   $\omega_2$ is unknown and maybe very small.  Fortunately, vectors come from $C_1^{\bot}$ do not affect the computation of $\omega_2$ because
they are degenerate.   For an integer $1<\lambda_X<n_1$, let $\nu\in GF(q)^{n_1}$ be any nonzero vector with Hamming weight less than  $ \lambda_{{X}}$. Denote  by
 \begin{equation}
 \textrm{S}_{\nu}\equiv G_1\nu^T .
 \end{equation}
If $\textrm{S}_{\nu}=0$, then $\nu$ must belong to $C_1^\bot$ and
$\nu$ is degenerate when we compute $\omega_2$. Therefore we only need to consider the case which makes $\textrm{S}_{\nu}\neq0$. Let $1\leq f_{X} \leq k_1$ and denote
by $k_{f_{{X}}}=k_1-(k_1-f_{{X}})/m$, where $m\geq 2$ is an integer and $m|(n_1-f_{{X}}) $.

  For some given $\mathbf{a}=(\alpha_1, \alpha_2, \ldots, \alpha_{k_1})\in GF(q^m)^{k_1}$ and varying $\mathbf{v}=(v_1, v_2, \ldots, v_n)\in GF(q^m)^{k_1}$, let $\textrm{GRS}_{k_{f_{{X}}}}(\mathbf{a}, \mathbf{v})=[k_1,{k_{f_{{X}}}},k_1-{k_{f_{{X}}}}+1]_{q^m}$ be a GRS code  over $GF(q^m)$. Let \begin{equation}
C_2\equiv\mathcal{A}_{k_1-k_{f_X}}(\textbf{a},\textbf{y})=\textrm{GRS}_{k_{f_{{X}}}}(\mathbf{a}, \mathbf{v})| GF(q)
\end{equation}
  be an alternant code, where $\mathbf{y}=(y_1, y_2, \ldots, y_{k_1})$ and $y_i\cdot v_i=1/\prod_{j\neq i}(\alpha_i-\alpha_j)$, for $1\leq i\leq {k_1}$. By Lemma \ref{dimanddisofalternantcodes},  the dimension of $C_2$ satisfies
  \begin{equation}
  k_2\geq k_1-m(k_1-k_{f_{{X}}})=f_{{X}}.
  \end{equation}
By Lemma \ref{numberofalternantcodes}, we know that the number
of alternant codes   that contain  $\textrm{S}_{\nu}$ does not relate to the Hamming weight of $\textrm{S}_{\nu}$ and is at most
\begin{equation}
(q^m-1)^{k_1-(k_1-f_{{X}})/m}.
\end{equation}
On the other hand, the total number of alternant codes $C_2$ is equal to the number of the choice of  $\mathbf{v}$, and it is equal to $(q^m-1)^{k_1}$ \cite{macwilliams1981theory}. According to the counting argument in \cite{macwilliams1981theory}, if
 \begin{equation}
 \label{QSCPC_Counting}
      (q^m-1)^{k_1-(k_1-f_X)/m}       \sum\limits_{j=1}^{ \lambda_X-1}(q-1)^j\binom{n_1}{j} <(q^m-1)^{ k_1 },
 \end{equation}
then there exists  an alternant code   such that there does not exist any  nonzero vector    $\nu  \in \mathscr{C}_{X}\backslash \mathscr{C}_Z^{\bot} $ with weight less than $\lambda_X$.
According to the estimate of Hamming ball in Lemma \ref{HammingBallEst} and taking  the limit as $n_1\rightarrow \infty$ in (\ref{QSCPC_Counting}), we can derive 
\begin{equation}
\label{GVC2}
 \frac{k_2}{n_1}\geq \frac{f_{{X}}}{ n_1}=\frac{k_1}{n_1}-H_q(\frac{\lambda_{{X}}}{n_1})-O(1).
\end{equation}

Combining (\ref{distanceQ1}), (\ref{GVC1}), and (\ref{GVC2})   above together, we have
 \begin{equation}
\left\{
             \begin{aligned}
              \frac{k_1}{n_1}&\geq 1-H_q(\frac{d_1}{n_1})-O(1),  \\
             \frac{k_2}{n_1}&\geq\frac{k_1}{n_1}-H_q(\frac{\lambda_{{X}}}{n_1})-O(1),\\
             d_{\mathscr{Q}}&\geq \min\{d_1,\lambda_{{X}}\}.
             \end{aligned}
\right.
\end{equation}
If we let $\lambda_{{X}}=d_1+c$, where $c$ is   any constant, then we have
\begin{equation}
\frac{k_{\mathscr{Q}}}{n_1}=\frac{k_2}{n_1} \geq1-2H_q(\frac{d_{\mathscr{Q}}}{n_1})-O(1).
\end{equation}

 \end{IEEEproof}

Suppose that we consider  an asymmetry between   $\omega_1$ and $\omega_2$ in the proof of Theorem \ref{Degenerate QSCPC}, and we apply the CSS construction for AQCs in Lemma \ref{AQC Constructions} to    PC-CSS codes. Then   asymmetric PC-CSS codes can attain the quantum GV bound for AQCs asymptotically.
\begin{corollarys}
\label{asymmetricenlarge}
There   exist asymptotically good asymmetric PC-CSS codes $\mathscr{A}=[[n_\mathscr{A},k_\mathscr{A},d_{Z}/d_{X}]]_q $ such that
\begin{equation}
 \frac{k_\mathscr{A}}{n_\mathscr{A}}\geq1-H_q(\delta_{X})-H_q(\delta_{Z})-O(1)
\end{equation}
for any block length $n_\mathscr{A}\rightarrow\infty$, where $\delta_{X}=d_{X}/n_\mathscr{A}$ and $\delta_{Z}=d_{Z}/n_\mathscr{A}$ are the relative  distances, and $0\leq\delta_{X}\leq\delta_{Z}\leq 1-1/q$.
\end{corollarys}

 It is known that   there exist families of  asymptotically good LDPC codes  which can achieve the classical GV bound. But the  dual of LDPC codes
 has a low-density generator matrix   and thus   the  minimum distance is  constant. If we use LDPC codes as the inner    code $C_1$, then
  the   minimum distance of   the PC-CSS code  without considering error degeneracy is   constant to the block length.
While  considering    error degeneracy,  the
  low weight codewords in the dual of LDPC codes do not affect the counting argument  in the proof of Theorem \ref{Degenerate QSCPC}.  Therefore, the PC-CSS code is extremely
  degenerate and the    minimum distance   is greatly improved from a constant to be linear with the block length.
 By applying the Evra-Kaufman-Z\'emor  distance balancing
  construction  \cite{evra2020decodable}, we show
  that our scheme can also lead to qLDPC codes wth non-vanishing   rates   and  minimum   distance  growing with the square root of the block length \cite{fan2021generalization}.

In    Corollary \ref{Degenerate QSCPC},  PC-CSS codes are shown to attain the   quantum GV bound for AQCs   \cite{matsumoto2017two}.
 Moreover, we show that  asymmetric PC-CSS codes can approach the   capacity of asymmetric Pauli channels as the channel asymmetry goes to large.
  Suppose that we transmit quantum information over the Pauli channel:
   \begin{equation}
   \varrho\mapsto p_I\varrho+p_X{X}\varrho {X}+p_Y {Y}\varrho  {Y}+p_Z{Z}\varrho {Z}
   \end{equation}
   for an input state $\varrho$, where $\{I,{X}, {Y},{Z}\}$ are the Pauli operators, $0\leq p_I,p_Z,p_Y,p_Z\leq 1$, and $p_I+p_X+p_Y+p_Z=1$.
   Denote  the total error probability by $\emph{\textbf{p}}=p_X+p_Y+p_Z$. Denote by $ \zeta=(p_Z+p_Y)/(p_X+p_Y)$ the asymmetry
   for the probabilities of ${Z}$-errors and ${X}$-errors. Usually, we let $p_X=p_Y$, then we have
   $p_X=\emph{\textbf{p}}/(2\zeta+1)$ and $p_Z=\emph{\textbf{p}}(2\zeta-1)/(2\zeta+1)$.

  In Corollary \ref{Degenerate QSCPC},   the inner code  $C_1$   can be chosen arbitrarily, we let it be an LDPC code, which is used to correct   $Z$-errors. We use   alternant codes to   correct   $X$-errors.  It is known that alternant codes can be decoded up to the GV bound under the bounded minimum distance decoder   \cite{macwilliams1981theory}.  LDPC codes can approach the Shannon capacity of the binary symmetric channel under the BP decoding \cite{mackay2003information,richardson2001capacity}. Then the rate $R_{\mathscr{Q}} $ of  PC-CSS codes   under the asymmetric Pauli channel can approach \begin{equation}
\label{hashingQSCPC}
   1-H_2(4p_X)-H_2(p_Z+p_Y)-O(1).
\end{equation}

\begin{figure}
\centering
\includegraphics[width=3.1in]{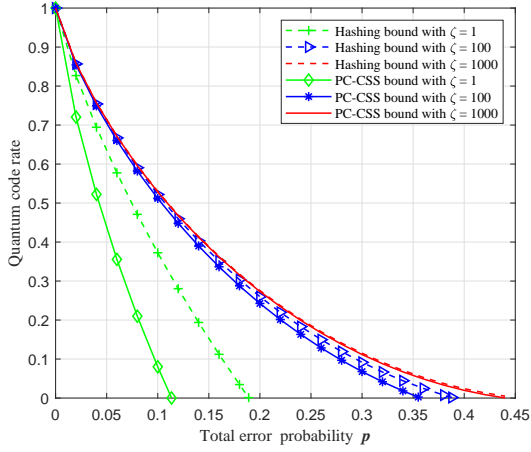}
\caption{The comparison of the asymptotic bound of asymmetric PC-CSS codes (solid lines) and the Hashing bound (dashed lines) with different channel asymmetries $\zeta=1,100,$ and $1000$. The horizontal axis is the total error probability in the Pauli channel  and the vertical axis is the quantum code rate.}\label{Compari1}
\end{figure}

 In Fig. \ref{Compari1}, we compare the the   limit  of PC-CSS codes in (\ref{hashingQSCPC}) with the Hashing bound given by $
R =1-H_2( \emph{\textbf{p}})
$  over Pauli channels with an asymmetry $ \zeta=(p_Z+p_Y)/(p_X+p_Y)$.
  If $\zeta=1$, then the channel is  symmetric with $p_X=p_Y=p_Z=\emph{\textbf{p}}/3$. As the asymmetry $\zeta$ grows,
  the gap between PC-CSS codes and the Hashing bound becomes increasingly smaller. In Fig. \ref{Compari1}, when the asymmetry $\zeta=10^2$ and $\zeta=10^{3}$, the code rate
  gaps are less than $3\times 10^{-2}$ and $4\times 10^{-3}$, respectively. Therefore, PC-CSS codes can approach the Hashing bound of Pauli channels as the asymmetry grows. It should be noted that, although our codes are highly degenerate, they can not attain the bound in \cite{PhysRevA57830} for very noisy channels  .

 The PC-CSS codes in Theorem \ref{Degenerate QSCPC} can also be used to prove the security of the famous BB84 quantum key distribution (QKD) protocol directly \cite{shor2000simple}. By a simple calculation, we can obtain a total qubit error rate (QBER) less than $7.56\%$ for  the security
 of BB84  over Pauli  channels.  Although the rate is lower than the  bound of $11\%$ in \cite{shor2000simple},  our codes are not randomly  generated and the LDPC component codes    can be efficiently decoded.   Since
 classical LDPC codes  can approach the Shannon bound by using efficient  BP decoders,    PC-CSS codes are particularly applicable to efficient and practical   QKD. Furthermore, if we consider asymmetric errors in Pauli channels, we can further
 improve the total QBER for the security of BB84. For example, if we admit an asymmetry $\zeta= 100$ between ${Z}$-errors and ${X}$-errors, then the total QBER is  less than $35.56\%$. Our codes are competitive with  quantum Polar
 codes in \cite{renes2012efficient,renes2015efficient} that can achieve the coherent information.  But Ref. \cite{renes2012efficient} needs some preshared entanglement assistance.

 \section{A Family of Fast Encodable and Decodable PC-CSS Codes}\label{FastDecEnc}
In practice, we need quantum codes to be   efficiently encoded and decoded. Let the error syndrome as the input, we show that PC-CSS codes can be fast decoded. Moreover, we also show that  PC-CSS codes   can be fast encoded.
\begin{table*}
\centering
 \caption{Parameters and  the syndrome decoding complexity of several efficiently decodable quantum codes.}
\begin{tabular}{|c|c|c|c|c|c| }
\hline
   References &Parameters    &   \makecell[c]{Weight of \\ $X$-Stabilizers }    &\makecell[c]{Weight of\\  $Z$-Stabilizers}  &  \makecell[c]{Syndrome\\Decoding Complexity}&\makecell[c]{Parallel Syndrome\\Decoding Complexity}   \\
\hline
   \cite{leverrier2015quantum} &  $[[N,\Omega(N),\Omega(\sqrt{N})]]$ &$ O(1)$ &$O(1)$&$O(N)$&$\backslash $\\
\hline
    \cite{Hamada2008concatenated} &$[[N,\Omega(N),\Omega( {N})]]$&$ O(N)$&$ O(N)$&Polynomial &$\backslash$\\
\hline
     \cite{kitaev2003fault} &$[[N,2,\Omega( \sqrt{N})]]$&$ O(1)$&$ O(1)$&$O(N^3)$ &$\backslash$\\
\hline
    \cite{duclos2010fast} &$[[N,2,\Omega( \sqrt{N})]]$&$ O(1)$&$ O(1)$&$O(N\log\sqrt{N})$ &$O(\log \sqrt{N})$\\
\hline
     \cite{evra2020decodable} &$[[N,\Omega(\sqrt{N/\log N}),\Omega(\sqrt{N\log N})]]$&$ O(1)$&$ O(1)$&Polynomial &$\backslash$\\
\hline
Corollary 2    & $[[N,\Omega(\sqrt{N}),\Omega(\sqrt{N})]]$ &$ O(\sqrt{N})$&$ O(1)$&$O( {N}) $&$O(\sqrt{N})$\\
\hline
Corollary 3  & \makecell[c]{$[[N,\Omega(N/\log N),d_Z/d_X]]$\\ $d_Z=\Omega(N/\log N),$\\ $d_X=\Omega(\log N)$}&$ O(\log N)$&$ O(1)$&$O(  N  ) $&$O(\log{N})$\\
\hline
\end{tabular}
\label{CompTable11}
\end{table*}

\begin{figure}
\centering
\hspace{6mm}\Qcircuit @C=0.5em @R=0.4em  {
   \lstick{\ket{0}}& \qw& \qw & \qw&\qw&\qw& \gate{H}&\ctrl{9}&\qw&\qw&\qw&\qw&\qw&\qw&\qw&\qw&\qw&\qw&\qw&\qw&\qw\\
   \lstick{}&&&\vdots& &&& &\ddots \\
   \lstick{\ket{0}}& \qw& \qw & \qw& \qw& \qw&\gate{H}&\qw&\qw&\qw&\ctrl{15}&\qw&\qw&\qw&\qw&\qw&\qw&\qw&\qw&\qw&\qw \\
      \lstick{}&&& &&& &&&& &&&  \\
      \lstick{}&&&\vdots&&& &&   &&&&\ddots\\
            \lstick{}&&& &&& &&&& &  \\
    \lstick{\ket{0}}& \qw& \qw & \qw& \qw&\qw&\gate{H}&\qw&\qw&\qw&\qw&\qw&\qw&\qw&\ctrl{3}&\qw&\qw&\qw&\qw&\qw&\qw\\
   \lstick{}&&&\vdots&&& &&   &&&& &&&&\ddots\\
   \lstick{\ket{0}}& \qw& \qw & \qw& \qw&\qw&\gate{H}&\qw&\qw&\qw&\qw&\qw&\qw&\qw&\qw&\qw&\qw& \qw&\ctrl{9}&\qw&\qw
      \\
    \lstick{\ket{0}}& \qw&\multigate{2}{\textrm{P}_{1}} &\qw& \multigate{2}{\textrm{P}_{r_2}}    &\qw&\qw &\targ&\qw&\qw&\qw&\qw&\qw&\qw&\targ&\qw&\qw&\qw&\qw&\qw&\qw  \\
            \lstick{}&&  &\cdots &          &&&&  &&&&&&&&&&&&&&&\\
    \lstick{\ket{0}}& \qw& \ghost{\textrm{P}_{1}} &\qw &\ghost{\textrm{P}_{r_2}}     &\qw& \qw&\qw &\qw&\qw&\qw&\qw&\qw&\qw&\qw&\qw&\qw&\qw&\qw&\qw&\qw \\
               \lstick{}  & &&&  &&&&&& &&&&&&&&\\
        \lstick{}  & \qw & \ctrl{-2}    &\qw & \qw  & \qw   & \qw&\qw&\qw& \qw &\qw&\qw&\qw&\qw&\qw&\qw&\qw&\qw&\qw&\qw&\qw
      \\
       \lstick{}  &  &&\ddots&&& &&& &&      & &&&&&&&&& \\
           \lstick{}  & &&&  &&&&&& &&&&&&&&\\
                             \lstick{}  & &&&  &&&&&& &&&&&&&&&&&\\
        \lstick{}   & \qw  &\qw   &\qw &  \ctrl{-6} & \qw& \qw& \qw&\qw&\qw&\targ&\qw&\qw&\qw&\qw&\qw&\qw&\qw&\targ&\qw&\qw
     \inputgroup{12}{12}{1.9em}{\ket{\psi}}   \\
                  \lstick{}  & &&&  &&&&&& &&&&&&&&&&&  \\
        \lstick{}  & &&\mbox{Stage I}&&&&&&&&&\mbox{Stage II}&&&&&&&&&&&
}
  \caption{The   encoding circuit of a family of PC-CSS codes. The box indicated by $\textrm{P}_i (1\leq i\leq r_2)$ means that the qubits pass it are controlled by a   qubit in quantum state $|\psi\rangle$ and if there is a ``1'' in the $j$th position of the $i$th column of the check part $P_2$ in $G_2$, then there is a C-NOT target ``$\otimes$'' in the $j$th qubit of the box $\textrm{P}_i$. }\label{EncodingCircuitBacon}
\end{figure}
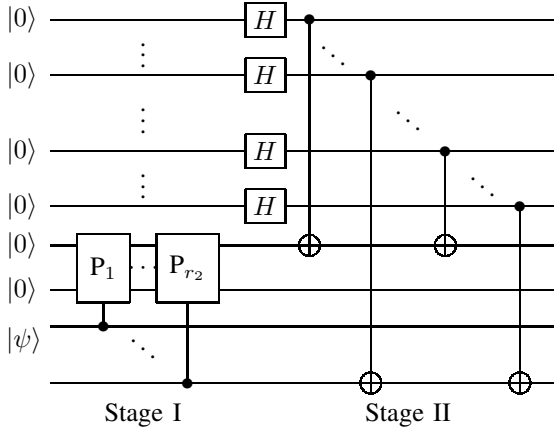

 \begin{theorem}
 \label{QSCPCdec1}
Let $n_0>1$ and $N>1$ be   integers and let $n_0|N$. There exists a family of asymmetric PC-CSS codes with parameters $\mathscr{A}=[[N,\Omega(N/n_0),d_{Z}\geq\Omega(N/n_0)/d_{X}\geq n_0]]$. The syndrome decoding of $\mathscr{A}$  can be done in $ O( N) $ time. Further, the  parallel syndrome decoding  of $\mathscr{A}$ can   be done in $ O(\log(N/n_0)+n_0) $  time  by using $O(N/n_0+N/n_0\cdot\log(N/n_0))$ classical processors.  Moreover, $\mathscr{A}$ can be encoded by a circuit of size $O((N/n_0)^2+N)$ and   depth   $ O(N/n_0+n_0)$.
 \end{theorem}
\begin{IEEEproof}
Let $C_0=[n_0,1,n_0]$ be a binary repetition code whose    parity check matrix  and   generator matrix  are given by:
\begin{equation}
H_0= \left(\begin{array}{ccccc}
  1&0&\cdots&0&1  \\
  0&1&\cdots&0&1  \\
  \vdots&\vdots&\ddots&\vdots&\vdots\\
  0&0&\cdots&1&1
 \end{array}\right),\ \textrm{and}\ G_0=  (\begin{array}{ccccc}
1&1&\cdots &1
 \end{array} ),
\end{equation}
respectively.
  Let $C_1= [N,N/n_0,n_0]$ be a linear code with  a parity check matrix $H_1=\mathbf{I}\otimes H_0$ and a generator matrix $G_1=\mathbf{I}\otimes G_0$, where $\mathbf{I}$ is an identity matrix of size $N/n_0$.
Let $C_2=[N_2=N/n_0,k_2,d_2]$ be an  asymptotically good expander code in \cite{sipser1996expander} such that
 $k_2=\gamma_2N_2$, and $d_2=\delta_2N_2$, where $0<\gamma_2,\delta_2<1$. Let the parity check and generator matrices of
$C_2$ be $H_2$ and $G_2$, respectively. Then we can construct an asymmetric PC-CSS code $\mathscr{A}= [[N,k_2,d_{Z}/d_{X}]]$ with
\begin{equation}
\mathscr{H}_{X} = H_{2}G_{1},\ \textrm{and}\  \mathscr{H}_{Z} = H_{1}.
\end{equation}
It is not difficult to verify that $d_{X}\geq d_2$ and $d_{Z}\geq n_0$.

We consider the syndrome decoding  of $\mathscr{A}$. For an  $X$-error $e_{X}$,  the error syndrome  is given by
\begin{equation}
\textrm{S}_X\equiv\mathscr{H}_X e_X^T= H_{2}\textrm{S}_X^{(1)},
\end{equation}
where we denote by $\textrm{S}_X^{(1)}=G_{1}e_X^T$. Suppose that  $\textrm{wt}_H(e_X)\leq \lfloor (d_2-1) /2 \rfloor$,
then we must have  $\textrm{wt}_H(\textrm{S}_X^{(1)})\leq \lfloor (d_2-1) /2 \rfloor$. Thus we can   decode
 $\textrm{S}_X^{(1)}=(\textrm{S}_{X_1}^{(1)},\ldots,\textrm{S}_{X_{N_2}}^{(1)})$ by using the expander code $C_2$ in $O(N_2)$ time \cite{sipser1996expander}. Further, $C_2$ can also be decoded in parallel in $O(\log(N_2))$ time  by using $O(N_2\cdot\log(N_2))$ classical processors \cite{sipser1996expander}. Let
  $\widetilde{e}_{X}=(\textrm{S}_{X_1}^{(1)},0,\ldots,\textrm{S}_{X_{N_2}}^{(1)},\ldots,0)$ be the decoded $X$-error, where the $0$s correspond to the redundant qubit positions in $G_1$. Then we have $G_{1}(e_X+\widetilde{e}_X)^T=0$ and $\mathscr{Q}$ is degenerate with respect to $e_X+\widetilde{e}_X$.

 Next, we use $C_1$ to correct any $Z$-error $e_Z$ such that
 $\textrm{wt}_H(e_Z)\leq \lfloor (n_0-1) /2\rfloor$. We divide  the parity check matrix $H_1$ into $n_0$
 sub-blocks by   columns, and each sub-block corresponds to a diagonal block $H_0$ in $H_1$. Divide  $e_Z$ into $N_2$ sub-blocks, i.e., $e_Z=(e_{Z_1},\ldots,e_{Z_{N_2}})$. Then there are at most
 $\lfloor (n_0-1) /2\rfloor$ erroneous sub-blocks in $e_Z$ and each erroneous sub-block has at most   $\lfloor (n_0-1) /2\rfloor$ errors.
  For each erroneous sub-block, we   use the one-step majority-logic (OSMLG) decoding  method \cite{lin2004error}  to check   the  error qubit that corresponds to the last column
 in $H_0$. Let   $\widehat{e}_Z=(\widehat{e}_{Z_1},\ldots,\widehat{e}_{Z_{n_0}})$ be any sub-block in $\{ e_{Z_i}|1\leq i\leq N_2\} $.   Then we can derive the following syndrome for $\widehat{e}_Z$:
 \begin{equation}
 \label{OSMLGSyn}
 S_{\widehat{e}_Z}=\widehat{e}_Z H_0^T=(\widehat{e}_{Z_1},\ldots,\widehat{e}_{Z_{n_0-1}})+\widehat{e}_{Z_{n_0}}E_0,
 \end{equation}
 where $E_0=(1,\ldots,1)$ is an all-ones vector of length $n_0-1$. If $\widehat{e}_{Z_{n_0}}=0$, then there are at most $\lfloor (n_0-1) /2\rfloor$ ones in the vector $(\widehat{e}_{Z_1},\ldots,\widehat{e}_{Z_{n_0-1}})$. Thus the Hamming weight of the syndrome $S_{\widehat{e}_Z}$ is at most $\lfloor (n_0-1) /2\rfloor$. While if $\widehat{e}_{Z_{n_0}}=1$, then there are at most $\lfloor (n_0-1) /2\rfloor-1$ ones in the vector $(\widehat{e}_{Z_1},\ldots,\widehat{e}_{Z_{n_0-1}})$. Then the Hamming weight of the syndrome $S_{\widehat{e}_Z}$ is at least $n_0-\lfloor (n_0-1) /2\rfloor=\lfloor (n_0-1) /2\rfloor+1$. Therefore we can use the OSMLG decoding to determine $\widehat{e}_{Z_{n_0}}$, and then $(\widehat{e}_{Z_1},\ldots,\widehat{e}_{Z_{n_0-1}})$  can be derived from (35) directly.

  The running time for deriving $\widehat{e}_Z=(\widehat{e}_{Z_1},\ldots,\widehat{e}_{Z_{n_0}})$  is determined by the computation of   the Hamming weight of the syndrome  $S_{\widehat{e}_Z}$.  The computation  time complexity is $O(n_0)$. Thus the running  time for decoding all the erroneous sub-blocks is  $O(N)$ since there are $N/n_0$ sub-blocks. Here, the decoding of the $n_0$ sub-blocks can also be carried out in parallel by using $N/n_0$ classical processors. Thus the parallel  syndrome decoding of the $Z$-error $e_Z$ can be done in $O(n_0)$ time by using  $N/n_0$ classical processors.    Overall, the  syndrome decoding of $\mathscr{A}$ can be done in $ O(N/n_0+N) $ time or can be done in parallel in $O(\log(N/n_0)+n_0) $ time  by using $O(N/n_0\cdot\log(N/n_0)+N/n_0)$ classical processors.

 Next, we consider the encoding of $\mathscr{A}$ and we use the standard encoding method of  CSS codes in \cite{cleve1997efficient}  to do that. We can encode the  pure state as follows:
 \begin{equation}
\label{AQC encoding}
|u +\mathscr{C}_Z^\bot\rangle   \equiv   \frac{1}{\sqrt{|\mathscr{C}_Z^\bot|}}\sum_{v \in \mathscr{C}_Z^\bot}|u +v \rangle,
\end{equation}
 where $u\in \mathscr{C}_X \backslash \mathscr{C}_Z^\bot $. It is not difficult to verify that the generator matrix of $\mathscr{C}_X $ is given by
 \begin{equation}
 \mathscr{G}_X=\left(
\begin{array}{cccc}
 H_1 \\
\mathbf{O}\ G_2
\end{array}
\right),
 \end{equation}
  where     $\mathbf{O}$ is a zero matrix of size $k_2\times (N-N/n_0)$, and $G_2=[I\ P_2]$ is in the standard form. Denote by $P_2=[ \textbf{P}_1,\ldots,  \textbf{P}_{r_2}]$, where each $ \textbf{P}_i(1\leq i\leq r_2)$ is a column of $P_2$ and $r_2=N/n_0-k_2$. Then the encoding circuit of the PC-CSS code $\mathscr{A}$ is given in Fig. \ref{EncodingCircuitBacon}. In Stage I of the encoding circuit, the number of gates is determined by the density of $P_2$ and it is at most $O((N/n_0)^2)$. It is easy to see that the depth of Stage I is $k_2$. While in Stage II, the number of C-NOT gates is exactly $N-N/n_0$ and the depth is $n_0$. Therefore the   size of the encoding circuit is $O((N/n_0)^2+N)$ and the depth is $k_2+n_0=O(N/n_0+n_0)$.
  \end{IEEEproof}

Notice that the  last column in $H_0$ in the proof of Theorem \ref{QSCPCdec1} is quite dense and then $H_1$ is also dense.  However, we can transform $H_0$ into a sparse matrix
 by multiplying a series of elementary matrices  in the left. Therefore   the ${Z}$-stabilizer generators of $\mathscr{A}$ satisfy the
  qLDPC constraint  \cite{hastings2021quantum} . In the fault tolerant settings, we require   quantum codes to correct a linear number of random errors. We consider the correction of random errors by using PC-CSS codes. We assume that the noise model used is the
 independent   error model
 in \cite{fawzi2018efficient,gottesman2014fault,dumer2015thresholds}.  Denote by $p_z$ the  probability of each  $Z$-error.
 It is shown that the PC-CSS code $\mathscr{A}$ in Theorem \ref{QSCPCdec1} can correct any
adversarial $Z$-error $e_Z$ of weight less than
half the $Z$ distance bound.
We  count all the uncorrectable errors of weight larger than half the minimum distance bound. It is easy to see that if there is at least one
sub-block that has $Z$-errors of weight larger than $d_0=\lfloor (n_0-1) /2\rfloor$, then $e_Z$ is undetectable. All the uncorrectable
$Z$-errors are counted as follows in equations (\ref{pz1}) - (\ref{pz3}).
\begin{figure*}[h]
 \hrulefill
 \begin{eqnarray} \label{pz1}
\mathcal{P}_{Z}&\leq& C_{N_2}^1C_{n_0}^{d_0+1} ( p_z^{d_0+1} (1-p_z)^{N-d_0-1}+C_{N-d_0-1}^1 p_z^{d_0+2} (1-p_z)^{N-d_0-2} +\cdots+C_{N-d_0-1}^{N-d_0-1} p_z^{N}  ) \\ \label{pz2}
&=& C_{N_2}^1C_{n_0}^{d_0+1}p_z^{d_0+1}  \sum_{i=0}^{N-d_0-1}C_{N-d_0-1}^{i }p_z^{ i}(1-p_z)^{N -d_0-1-i}\\ \label{pz3}
&\leq& N_2 2^{n_0-1} p_z^{d_0+1}  ,
 \end{eqnarray}
\hrulefill
\end{figure*}

We denote by
 $\mathcal{P}_Z$   the probability of uncorrectable $Z$-errors and denote by  $  N_2=N/n_0$. Let  $c>0$ be any constant. If we let $n_0=\Omega(\log N)$ and let $ p_z<1/4^{c+1} $, then $N^c\mathcal{P}_Z$ vanishes as $N\rightarrow\infty$. If we let $n_0=  \Theta( N^{c_0}) $, where $0<c_0<1$ is  constant, then $N^c\mathcal{P}_Z$   vanishes when $p_z<25\%$ as $N\rightarrow\infty$. Therefore the PC-CSS code $\mathscr{A}$ in Theorem \ref{QSCPCdec1} can correct a linear number of random $Z$-errors with high probability in $O(n_0)$ time as long as $n_0=\Omega(\log N)$. However, for the correction of random ${Z}$-errors, we cannot get a non vanishing threshold.

If  we let $n_0=\Theta(\sqrt{N})$ or $n_0=\Theta(\log N)$,     we can derive the following two families of fast encodable and decodable quantum codes.
\begin{corollarys}\label{fastcode1}
There exists
  a family of quantum codes with parameters
$
  \mathscr{Q}_1=[[N,\Omega(\sqrt{N}),\Omega(\sqrt{N})]].
$
  For an input error syndrome,  $\mathscr{Q}_1$   can correct all errors of weight smaller than  half   the minimum distance bound   in  $O({N})$ time.  $\mathscr{Q}_1$ can also be decoded   in parallel in $O(\sqrt{N})$ time   by using $ O(\sqrt{N})$ classical processors.  We show that $\mathscr{Q}_1$ can be encoded efficiently by a circuit of size $O(N)$ and depth $O(\sqrt{N})$. Moreover, $\mathscr{Q}_1$ can correct a random $Z$-error   with very high probability  provided $p_z<25\%$.
\end{corollarys}

\begin{corollarys}\label{fastcode2}
There exists a family of  quantum codes with parameters
$
\mathscr{Q}_2=[[N,\Omega(N/\log N),\Omega(N/\log N)/\Omega(\log N)]].
$
 For an input error syndrome, $\mathscr{Q}_2$ can correct an almost linear number  of  ${Z}$-errors   and can also correct a linear number of ${X}$-errors with   high probability in  $O(N )$ time. Further, $\mathscr{Q}_2$ can be decoded in parallel in $O(\log(N))$ time   by using $O(N)$ classical processors.
\end{corollarys}

 In \cite{Hamada2008concatenated}, families of concatenated quantum AG codes were constructed and decoded efficiently in
polynomial time, while our codes in Theorem \ref{QSCPCdec1} can be decoded in linear time.  Therefore  our codes are more efficient than   concatenated quantum AG codes in \cite{Hamada2008concatenated}. In particular, the parallel syndrome decoding of  Corollary \ref{fastcode1} can be carried out in sublinear  time, and that of   Corollary \ref{fastcode2} can be carried out  in logarithmic time. In Table II, we compare the parameters and decoding complexity of several efficiently decodable quantum codes.

\section{Enlargement of PC-CSS Codes  Achieving  the Quantum GV Bound for Enlarged CSS Codes Asymptotically}
\label{enlargementSteane}
In \cite{Steane796388,Ling5508639}, an enlargement construction of CSS codes was proposed by enlarging  the dual-containing   codes. Further, several upper and lower bounds for the enlargement of CSS
codes were given in \cite{Steane796388,Ling5508639}. Among them, whether enlargement of CSS codes can attain the quantum GV bound for enlarged CSS codes is unknown. In this work, we show that
enlarged PC-CSS codes can attain the quantum GV bound asymptotically.
\begin{theorem}
\label{enlargedCSSGV}
There exist a family of enlarged stabilizer codes  with parameters $  \mathscr{Q}=[[n_\mathscr{Q},
k_\mathscr{Q},d_\mathscr{Q}]]_q$ achieving  the quantum GV bound for enlarged CSS codes asymptotically. $\mathscr{Q}$ can achieve a quantum rate
\begin{equation}
  \frac{k_\mathscr{Q}}{n_\mathscr{Q}}\geq1- H_{q}(\delta_\mathscr{Q})-H_{q}(\frac{q}{q+1}\delta_\mathscr{Q} )-O(1)
\end{equation}
for any block length $n_\mathscr{Q}\rightarrow\infty$, where $\delta_\mathscr{Q}=d_\mathscr{Q}/n_\mathscr{Q}$ is the relative minimum distance.
\end{theorem}
\begin{IEEEproof}
  Suppose that $C_1=[n_1,k_1,d_1]_q$ is a dual-containing code, i.e., $C_1^\bot\subseteq C_1$.  Denote the parity check and generator matrices of $C_1$ by $H_1$ and $G_1$, respectively. We begin with a family of   dual-containing codes. It is known that there exists   asymptotically good dual-containing codes $C_1^\bot\subseteq C_1$  achieving  the quantum GV bound \cite{ashikhmin2000quantum,ashikhmin2000quantum2}, i.e., \begin{equation}\label{EnlargedCSSGV1}
 \frac{k_1}{n_1}\geq1-H_q(\delta_1)-O(1),\ n_1\rightarrow\infty,
 \end{equation}
 where $\delta_1=d_1/n_1$ is the relative distance of $C_1$. Let $r_1=n_1-k_1$ and let $C_2=[r_1,k_2,d_2]_q$ be an alternant code. Denote the parity check and generator matrices of $C_2$ by $H_2$ and $G_2$, respectively. We construct a linear code $C_3=[n_1,k_1+k_2,d_3]_q$ with a parity check matrix given by
  \begin{equation}
  H_3=H_2H_1.
  \end{equation}
 Then there is $C_1^\bot\subseteq C_1\subseteq C_3$. According to the enlarged CSS construction in Lemma \ref{enlargedCSSConstruction}, we can derive a stabilizer code with parameters
$
 \mathscr{Q}=[[n_1,k_{\mathscr{Q}}=2k_1+k_2-n_1,d_{\mathscr{Q}}=\min\{d_1,\omega_2\}]]_q,
$
 where     $\omega_2= \lceil\frac{q+1}{q}d_3 \rceil  $.
   We need to determine $\omega_2$ and we chose $C_2=[r_1,k_2,d_2]_q$ as   an alternant code.

  Let $1\leq f_{A} \leq r_1$ and denote
by $k_{f_{A}}=r_1-(r_1-f_{A})/m$, where $m\geq 2$ is an integer and $m|(n_1-f_{A}) $.   For some given $\mathbf{b}=(\beta_1, \beta_2, \ldots, \beta_{r_1})\in GF(q^m)^{r_1}$ and varying $\mathbf{u}=(u_1, u_2, \ldots, u_{r_1})\in GF(q^m)^{r_1}$,   let $\textrm{GRS}_{k_{f_{A}}}(\mathbf{b}, \mathbf{u})=[r_1,{k_{f_{A}}},r_1-{k_{f_{A}}}+1]_{q^m}$ be a GRS code  over $GF(q^m)$. Let  \begin{equation}
C_2\equiv\mathcal{A}_{r_1-k_{f_{A}}}(\textbf{b},\textbf{z})=\textrm{GRS}_{k_{f_A}}(\mathbf{b}, \mathbf{u})| GF(q)
\end{equation}
  be an alternant code, where $\mathbf{z}=(z_1, z_2, \ldots, z_{r_1})$ and $z_i\cdot u_i=1/\prod_{j\neq i}(\beta_i-\beta_j)$, for $1\leq i\leq r_1$. By Lemma \ref{dimanddisofalternantcodes}, the dimension of $C_2$ satisfies
  \begin{equation}
  k_2\geq r_1-m(r_1-k_{f_A})=f_{A}.
  \end{equation}
  Let $ 1< \lambda_A < n_1$ be an integer. Let $\upsilon \in GF(q)^{n_1}$ be any nonzero vector  with Hamming weight less than  $ \lambda_A$.
   Denote by  $ \textrm{S}_{\upsilon}= H_1\upsilon^T$.  If $\lambda_A<d_1$, then we must have $ \textrm{S}_{\upsilon}\neq 0$. If $\upsilon\in C_3$, then we have $H_2\textrm{S}_{\upsilon}=H_2H_1\upsilon^T=0$. By Lemma \ref{numberofalternantcodes}, we know that the number
of alternant codes   that contain  the nonzero vector $\textrm{S}_{\upsilon}$ does not relate to the Hamming weight of $\textrm{S}_{\upsilon}$ and is at most
\begin{equation}
(q^m-1)^{r_1-(r_1-f_{A})/m}.
\end{equation}
On the other hand, the total number of alternant codes $C_2$ is equal to the number of the choice of  $\mathbf{u}$, and it is equal to $(q^m-1)^{r_1}$. According to the counting argument in \cite{macwilliams1981theory}, if
 \begin{equation}
 \label{enlargedCSSCounting2}
         (q^m-1)^{r_1-(r_1-f_{A})/m}    \sum\limits_{j=1}^{\lambda_{A}-1}(q-1)^j\binom{n_1}{j} <(q^m-1)^{ r_1 },
 \end{equation}
then there exists  an alternant code   such that there does not exist any  nonzero vector    $\upsilon\in C_3 $ with weight less than $\lambda_A$. Therefore we have $d_3\geq \lambda_{A}$.
 According to the estimate of Hamming ball in Lemma \ref{HammingBallEst}  and taking  the limit as $n_1\rightarrow \infty$ in (\ref{enlargedCSSCounting2}), we can derive 
\begin{equation}
\label{EnlargedCSSGV3}
  \frac{k_2}{n_1}\geq \frac{f_A}{ n_1}=\frac{r_1}{n_1}-H_q(\frac{\lambda_A}{n_1})-O(1).
\end{equation}

Combining (\ref{EnlargedCSSGV1}),    (\ref{EnlargedCSSGV3}), and the parameters of $ \mathscr{Q}$   above together, we have
 \begin{equation}
\left\{
             \begin{aligned}
              \frac{k_1}{n_1}&\geq 1-H_q(\frac{d_1}{n_1})-O(1),  \\
             \frac{k_2}{n_1}&\geq\frac{r_1}{n_1}-H_q(\frac{\lambda_A}{n_1})-O(1),\\
              d_{\mathscr{Q}}&\geq \min\{d_1,\lceil\frac{q+1}{q}\lambda_A\rceil\},\\
              d_1&>\lambda_A.
             \end{aligned}
\right.
\end{equation}
  Then there is
 \begin{equation}
\left\{
             \begin{aligned}
              \frac{k_{\mathscr{Q}}}{n_1}&\geq 1-H_q(\frac{d_1}{n_1})-H_q(\frac{\lambda_A}{n_1})-O(1),  \\
            d_{\mathscr{Q}}&\geq \min\{d_1,\lceil\frac{q+1}{q}\lambda_A\rceil\},\\
             d_1&>\lambda_A.
             \end{aligned}
\right.
\end{equation}
  If we let $d_1= \lceil\frac{q+1}{q}\lambda_A\rceil$, then we have
\begin{equation}
\frac{k_{\mathscr{Q}}}{n_1}  \geq1- H_q(\frac{d_{\mathscr{Q}}}{n_1})-H_q(\frac{q}{q+1}\cdot\frac{d_{\mathscr{Q}}}{n_1})-O(1).
\end{equation}

\end{IEEEproof}
\section{Discussions and Conclusions}
\label{DisCon}
 Constructing asymptotically good QECCs is one central problem  in quantum coding theory. In this paper, we have proposed a partial concatenation  construction of CSS codes by using   alternant codes as the outer code and any linear code achieving  the GV bound as the inner code. We   show that   PC-CSS codes can attain  the quantum GV bound for CSS codes and AQCs asymptotically. Further,  we   show  that PC-CSS codes can approach the Hashing bound for asymmetric Pauli channels with a large asymmetry.  In addition,   PC-CSS codes can   be used to prove the security of  the BB84  QKD  protocol. Moreover, the concatenation scheme  can asymptotically attain  the quantum GV bound for enlarged CSS codes by using Steane's enlargement construction of CSS codes.

 We   derive  a family of PC-CSS codes that can be encoded by a circuit of linear size and sublinear depth, and the syndrome decoding of PC-CSS codes can be done in linear time. Moreover, we   show  that the parallel syndrome decoding of the family of PC-CSS codes  can be done in sublinear time.   We     derive  a family of PC-CSS codes that can correct   a linear number of ${X}$-errors with    high probability and an almost linear number  of  ${Z}$-errors in linear time.  The parallel syndrome decoding of the family of PC-CSS codes can be done in logarithmic time.
One interesting future work is
 how to compute the threshold of PC-CSS codes for correcting random quantum errors.

\section*{Acknowledgment}
The authors would like  to thank  the anonymous referees and the Associate Editor  for their  valuable comments and helpful suggestions.

\bibliographystyle{IEEEtran}
\bibliography{IEEEabrv,IT-21-0825}

\begin{IEEEbiographynophoto}{Jihao Fan}
received the B.S.
degree in Mathematics and Applied Mathematics
from Lanzhou University, Lanzhou, China, in 2009,
and the Ph.D. degree in Computer Software and
Theory from Southeast University, Nanjing, China,
in 2016. He is currently an Associate  Professor with
the Nanjing University of Science and Technology,
Nanjing, China. His research interests include classical
and quantum coding theory, information theory, and
machine learning.
\end{IEEEbiographynophoto}

\begin{IEEEbiographynophoto}{Jun Li}
(M'09-SM'16)   received Ph. D degree in Electronic Engineering from Shanghai Jiao Tong University, Shanghai, P. R. China in 2009. From January 2009 to June 2009, he worked in the Department of Research and Innovation, Alcatel Lucent Shanghai Bell as a Research Scientist. From June 2009 to April 2012, he was a Postdoctoral Fellow at the School of Electrical Engineering and Telecommunications, the University of New South Wales, Australia. From April 2012 to June 2015, he was a Research Fellow at the School of Electrical Engineering, the University of Sydney, Australia. From June 2015 to now, he is a Professor at the School of Electronic and Optical Engineering, Nanjing University of Science and Technology, Nanjing, China. He was a visiting professor at Princeton University from 2018 to 2019. His research interests include network information theory, game theory, distributed intelligence, multiple agent reinforcement learning, and their applications in ultra-dense wireless networks, mobile edge computing, network privacy and security, and industrial Internet of things. He has co-authored more than 200 papers in IEEE journals and conferences, and holds 1 US patents and more than 10 Chinese patents in these areas. He is serving as an editor of IEEE Transactions on Wireless Communication and TPC member for several flagship IEEE conferences.
\end{IEEEbiographynophoto}

\begin{IEEEbiographynophoto}{Ya Wang}
 received the Ph.D. degree in Physics from the University of Science and Technology of China (USTC) in 2012. From August 2012 to July 2016, he was a Postdoctoral Fellow at Stuttgart University. From July 2016 to March 2018, he was a Research Fellow at the University of Science and Technology of China. From April 2018 to now, he has been a professor with the School of Physical Sciences of USTC. His research interests include spin-based quantum devices, spin quantum control, and their applications in quantum science. He currently focuses on diamond quantum device engineering and applications in quantum technologies.
\end{IEEEbiographynophoto}

\begin{IEEEbiographynophoto}{Yonghui Li}
 (M'04-SM'09-F'19)  received his PhD degree in November 2002 from Beijing University of Aeronautics and Astronautics. Since 2003, he has been with the Centre of Excellence in Telecommunications, the University of Sydney, Australia. He is now a Professor and Director of Wireless Engineering Laboratory in School of Electrical and Information Engineering, University of Sydney. He is the recipient of the Australian Queen Elizabeth II Fellowship in 2008 and the Australian Future Fellowship in 2012. He is a Fellow of IEEE.

His current research interests are in the area of wireless communications, with a particular focus on MIMO, millimeter wave communications, machine to machine communications, coding techniques and cooperative communications. He holds a number of patents granted and pending in these fields. He is now an editor for IEEE transactions on communications, IEEE transactions on vehicular technology. He also served as the guest editor for several IEEE journals, such as IEEE JSAC, IEEE Communications Magazine, IEEE IoT journal, IEEE Access. He received the best paper awards from IEEE International Conference on Communications (ICC) 2014, IEEE PIRMC 2017 and IEEE Wireless Days Conferences (WD) 2014.
\end{IEEEbiographynophoto}

\begin{IEEEbiographynophoto}{Min-Hsiu Hsieh}
received
the B.S. and M.S. degrees in electrical engineering
from National Taiwan University in 1999 and
2001, respectively, and the Ph.D. degree in electrical
engineering from the University of Southern
California at Los Angeles, Los Angeles, CA, USA,
in 2008. From 2008 to 2010, he was a Researcher
with the ERATO-SORST Quantum Computation and
Information Project, Japan Science and Technology
Agency, Tokyo, Japan. From 2010 to 2012, he was
a Post-Doctoral Researcher with the Statistical Laboratory,
the Centre for Mathematical Sciences, The University of Cambridge,
U.K. From 2012 to 2020, he was an Australian Research Council (ARC)
Future Fellow and an Associate Professor with the Centre for Quantum Software
and Information, Faculty of Engineering and Information Technology,
University of Technology Sydney, Ultimo, NSW, Australia. He is currently
the Director of the Hon Hai (Foxconn) Quantum Computing Center. His
scientific research interests include quantum information, quantum learning,
and quantum computation.
\end{IEEEbiographynophoto}

\begin{IEEEbiographynophoto}{Jiangfeng Du}
 received the Ph.D. degree in Physics from the University of Science and Technology of China (USTC) in 2000. Since 2004, he has been a professor with the School of Physical Sciences of USTC. He is an expert in the area of spin quantum physics and its applications. His research interests are interdisciplinary. He currently focuses on quantum computation, quantum simulation and quantum sensing, and the development of new technologies for medical applications.
\end{IEEEbiographynophoto}

\end{document}